\documentclass[
twocolumn,%
tightenlines,showpacs,%
nofootinbib,%
superscriptaddress,%
amsfonts,amsmath]{revtex4}

\usepackage{hyperref}
\usepackage{amsmath}
\usepackage{amssymb}
\usepackage{amsthm}

\begin{document}

\def\be{\begin{equation}}
\def\ee{\end{equation}}
\def\bea{\begin{eqnarray}}
\def\eea{\end{eqnarray}}
\def\ba{\begin{array}}
\def\ea{\end{array}}
\def\bse{\begin{subequations}}
\def\ese{\end{subequations}}

\def\pd{{\partial}}

\def\vp{{\varphi}}
\def\L{{\mathcal{L}}}
\def\E{{\mathcal{E}}}

\title{Plane waves in the generalized Galileon theory}

\author{Eugeny~Babichev} 
\affiliation{Laboratoire de Physique Th\'eorique d'Orsay,
B\^atiment 210, Universit\'e Paris-Sud 11,
F-91405 Orsay Cedex, France}
\affiliation{${\mathcal{G}}{\mathbb{R}}
\varepsilon{\mathbb{C}}{\mathcal{O}}$, Institut d'Astrophysique
de Paris, UMR 7095-CNRS, Universit\'e Pierre et Marie
Curie-Paris 6, 98bis boulevard Arago, F-75014 Paris, France}

\begin{abstract}
We present an exact plane wave solution of the most general shift-symmetric Horndeski  (generalized Galileon) theory.
The solution consists of the scalar part, and the gravitational part with two polarization modes.
The former is due to the presence of the non-trivial Galileon scalar field, and
it is parametrized by an arbitrary function of the light-cone coordinate.
For a trivial scalar field configuration the solution is equivalent to the plane gravitational wave in General Relativity.
When the metric is Minkowski,
we reproduce known results for the plane waves of
$k$-essence and a soliton-like solutions of a non-covariant Galileon model in a flat space-time.
\end{abstract}

\date{\today}

\pacs{04.50.Kd, 04.20.Jb, 11.10.Lm}

\maketitle

\section{Introduction}

The general scalar-tensor theory whose field equations for the metric as well as for the scalar field involve
at most second-order derivative was first formulated by Horndeski~\cite{Horndeski}.
A remarkable property of the theory
lies in the fact that in spite of its Lagrangian contains higher-order derivative terms, the associated field equations are of second-order.
This theory can be thought as a generalization of the
well-known scalar-tensor Brans-Dicke theory~\cite{Jordan} and its extensions~\cite{Damour:1992we}.

Later, in a quite different context a similar model, dubbed Galileon, was introduced in~\cite{Nicolis:2008in}.
The original Galileon is a scalar field theory, living in a flat space-time and invariant under the Galilean transformations of the field.
A covariant version of Galileon (``covariant Galileon'') was found in \cite{Deffayet:2009wt},
and further generalized in \cite{Deffayet:2011gz} (see also \cite{Deffayet:2009mn}).
It turns out that the most general covariant Galileon in four dimensions coincides with the Horndeski theory \cite{Kobayashi:2011nu}.

The Horndeski model is interesting in several aspects.
First of all, since  this theory contains non-quadratic kinetic interactions,
the perturbations propagate in an effective metric, which is in general different from the gravitational one.
In particular, perturbations may travel superluminally.
Another interesting feature of the Horndeski theory is the presence of the screening effect.
Since the Galileon model can be viewed as a certain $k$-mouflage theory, which
generically possesses the Vainshtein mechanism \cite{Babichev:2009ee},
one expects the same mechanism to be also present in the Galileon theory \cite{Kimura:2011dc}.
This property can also be seen from a different perspective:
the decoupling limit of the ghost-free massive gravity gives rise to (a part of) the Galileon Lagrangian,
while the Vainshtein mechanism has been shown to operate for massive gravity~\cite{Babichev:2009jt,Volkov:2012wp}.

Both superluminality and the screening in the Galileon theory are due to the non-linear kinetic mixing of the theory.
On the other hand, the non-linearity renders difficult to find exact solutions of the theory. 
It is not surprising that only a few exact solutions were found for (certain classes) of the Horndeski theory.
Therefore it is of interest to look for other exact solutions.

In the case when the metric is flat (and its dynamics is neglected) 
the Horndeski theory becomes a scalar field theory, which is easier to handle.
In this limit,  light-like waves of non-linear scalar field theories 
were presented in the literature before. 
In particular, in the context of the  $k$-essence model ---
which can be viewed as a subclass of the Galileon model --- such solutions were found in \cite{Babichev:2007dw}.
Also, a certain class of the non-covariant version of Galileon
model (on flat background)  possesses light-like waves~\cite{Evslin:2011vh} 
(for a more recent work see also \cite{Deser:2012gm}), 
which sometimes dubbed ``moving solitons'' or ``moving domain walls''.
The stability of these solutions 
were further studied in~\cite{Evslin:2011rj}.

In this paper we present an exact plain wave solution in the context of the most general shift symmetric Horndeski theory,
taking into account dynamics of the metric as well.
The theory that we study includes, among others, pure $k$-essence theories, 
the decoupling limit of the Dvali- Gabadadze-Porrati (DGP) model
and the covariantized Galileon.
In flat space-time the solution reproduces known light-like wave solutions
for the pure $k$-essence model,
and for the DGP-like non-covariant Galileon.

The paper is organized as follows. In Sec.~\ref{generalities} we give the general expressions of the action,
and the equations of motion for the scalar field and the metric.
In Sec.~\ref{solutions} we consider a $pp$-wave ansatz for the metric and we assume that the scalar field only depends on the retarded time coordinate. 
We will then exhibit a plane wave solution, and the last section is for the conclusions and the further directions to investigate in the future.

\section{Action, energy-momentum tensor and equations of motion} \label{generalities}

The most general shift-symmetric action, giving rise to the equations of motion up to the second order, can be parameterized by
four free (dimensionfull) functions of the standard kinetic term,
	\be
	K(X), \quad G^{(n)}(X),\; n=3,4,5,\nonumber
	\ee
where $X$ stands for the canonical kinetic term,
	\be
	X = -\frac12 g^{\mu\nu}\pd_\mu\vp \pd_\nu\vp.\nonumber
	\ee
The full action can be written as,
	\be
	\label{action}
		S = \int d^4 x \sum_{n=2}^5 \L_{n},
	\ee
The Galileon terms entering this action are,
    \begin{eqnarray}
        \mathcal{L}_{2} & = & K\left(X\right),	\label{L2}\\
        \mathcal{L}_3 & = & G^{(3)}\left(X\right)\Box\vp,\label{L3}\\
        \mathcal{L}_4 & = & G_{,X}^{(4)}(X)\left[\left(\Box\vp\right)^{2}-\left(\nabla\nabla\vp\right)^{2}\right]
        +R\, G^{(4)}(X), \label{L4} \\
        \mathcal{L}_5 & = & G_{,X}^{(5)}\left(X\right)\left[\left(\Box\vp\right)^{3}-3\Box\vp\left(\nabla\nabla\vp\right)^{2} +2\left(\nabla\nabla\vp\right)^{3}\right]
        \nonumber\\
        && -6G_{\mu\nu}\nabla^{\mu}\nabla^{\nu}\vp\,
        G^{(5)}\left(X\right), \label{L5}
    \end{eqnarray}
where the following short-hand notations are used
	\be
	\begin{aligned}
		\left(\nabla\nabla\vp\right)^{2} &= \left(\nabla_{\mu}\nabla_{\nu}\vp\right)\left(\nabla^{\mu}\nabla^{\nu}\vp\right),\nonumber\\
		\left(\nabla\nabla\vp\right)^{3} &=
		\left(\nabla_{\mu}\nabla_{\nu}\vp\right)\left(\nabla^{\mu}\nabla^{\rho}\vp\right)\left(\nabla_{\rho}\nabla^{\nu}\vp\right).\nonumber
	\end{aligned}
	\ee
The subscript of each Lagrangian term corresponds to the number of fields $\vp$, which enters this term, when
considered in flat space-time with $K(X) = G^{(3)} = X$ and $G^{(4)}=G^{(5)}  = X^2$.
These notations are in accordance with those of the original non-covariant version of the Galileon.
We have not included the Lagrangian $\L^{(1)}$, since it
would break the shift-symmetry. Note that the Einstein-Hilbert term can be absorbed in the last term of (\ref{L4}),
such that the action for General Relativity is recovered by identifying $G^{(4)} = M_P^2$, 
where $M_P$ is the reduced Planck mass.

Thanks to the shift symmetry, the equation of motion for the scalar field can be written in the form of a current conservation,
	\be
	\label{eomJ}
	\nabla_\mu J^\mu = 0,
	\ee
where the total current is the sum of the currents corresponding to each of the four Lagrangians,
	\be
		J_\mu = \sum_{n=2}^5 J^{(n)}_\mu.\nonumber
	\ee
The expression for the currents can be written as \cite{Gao:2011qe},
\begin{eqnarray}
	J^{(2)}_\mu & = & K_{,X}^{}\nabla_{\mu}\vp, \label{J2}\\
	J^{(3)}_\mu & = & \square\vp\, G_{,X}^{(3)}\nabla_{\mu}\vp+\nabla_{\mu}G^{(3)}, \label{J3} \\
	J^{(4)}_\mu & = & \left[\left(\left(\square\vp\right)^{2}-\left(\nabla\nabla\vp\right)^{2}\right)G_{,XX}^{(4)} + RG_{,X}^{(4)}\right]\nabla_{\mu}\vp \nonumber\\
		&+& 2\nabla_{\nu}\left(G_{,X}^{(4)}\left(\Box\vp\, \delta_\mu^\nu - \nabla_{\mu}\nabla^{\nu}\vp\right)\right), \label{J4}
\end{eqnarray}
and the expression for $J^{(5)}_\mu$ in given in the appendix~\ref{appendix}.
The energy-momentum tensor is also the sum of individual contributions,
\be
T_{\mu\nu} = \sum_{n=2}^5 T^{(n)}_{\mu\nu},\nonumber
\ee
where the $k$-essence part of the energy-momentum tensor is,
\be
        T_{\mu\nu}^{(2)}  =  K^{}g_{\mu\nu}+K_{,X}^{}\nabla_{\mu}\vp\nabla_{\nu}\vp. \label{T2}
\ee
The contribution from the generalized DGP (``kinetic gravity braiding'' \cite{Deffayet:2010qz}) term, $\L_3$, is
    \be
    \label{T3}
    \begin{aligned}
        T_{\mu\nu}^{(3)} & =  - \left(\nabla_{\lambda} G^{(3)}  \nabla^{\lambda}\vp \right) g_{\mu\nu} + 2 \nabla_{(\mu} G^{(3)} \nabla_{\nu)} \vp \\
        				&+ \square\vp G_{,X}^{(3)} \nabla_{\mu}\vp\nabla_{\nu}\vp
				          - \left(G_{,X}^{(3)}\nabla_{\lambda}X \nabla^{\lambda}\vp\right)g_{\mu\nu}
        	\\
        &  + \left(G_{,X}^{(3)}\square\vp\right)\nabla_{\mu}\vp\nabla_{\nu}\vp+2G_{,X}^{(3)}\nabla_{(\mu}\vp\nabla_{\nu)}X,
    \end{aligned}
     \ee
the higher-order $\L_4$ part reads,
\begin{widetext}
    \be
    \label{T4}
    \begin{aligned}
        T_{\mu\nu}^{(4)} & =  g_{\mu\nu}\Big\{ RG^{(4)}-G_{,X}^{(2)}\left(\left(\square\vp\right)^{2}-\left(\nabla\nabla\vp\right)^{2}\right) - 2G_{,XX}^{(2)}\nabla_{\rho}X				\nabla^{\rho}X  -2\left(\square\vp G_{,XX}^{(4)}\right)\nabla_{\rho}\vp\nabla^{\rho}X
	\\
         &   + 2G_{,X}^{(4)}R_{\rho\sigma}\nabla^{\rho}\vp\nabla^{\sigma}\vp\Big\}
          +\left[G_{,X}^{(4)}R +G_{,XX}^{(2)}\left(\left(\square\vp\right)^{2}-\left(\nabla\nabla\vp\right)^{2}\right)\right]\nabla_{\mu}\vp\nabla_{\nu}\vp
             +4\left(\square\vp G_{,XX}^{(4)}\right)\nabla_{(\mu}\vp\nabla_{\nu)}X
             \\
             &+2G_{,XX}^{(4)}\left(\nabla_{\mu}X\nabla_{\nu}X-2\nabla_{\rho}X\nabla^{\rho}\nabla_{(\mu}\vp\nabla_{\nu)}\vp\right)
              +2\left(G_{,X}^{(4)}\square\vp+G_{,XX}^{(4)}\nabla_{\rho}\vp\nabla^{\rho}X\right)\nabla_{\mu}\nabla_{\nu}\vp\\
         &   -2G_{,X}^{(4)}\left(\nabla^{\rho}\nabla_{\mu}\vp\nabla_{\nu}\nabla_{\rho}\vp+2\nabla_{(\mu}\vp R_{\nu)\rho}\nabla^{\rho}\vp+R_{\rho\mu\sigma\nu}\nabla^{\rho}\vp\nabla^{\sigma}\vp\right)-2G^{(4)}R_{\mu\nu},
        \end{aligned}
        \ee
\end{widetext}
and the expression for the energy-momentum tensor corresponding to $\L^{(5)}$ can be found in the appendix~\ref{appendix}.
The equations of motion, obtained by varying with respect to the metric (generalized Einstein equations) read,
\be
T_{\mu\nu} = 0. \label{eomg}
\ee
Note that if $\L_4$ contains the Einstein-Hilbert term, then the Einstein tensor appears as a part of $T^{(4)}_{\mu\nu}$ in the above equation.

\section{Ansatz and solution}\label{solutions}
%
After having written all the necessary expressions, let us consider the following ansatz for the metric,
	\be
		\label{metric}
		ds^2 = - F(u,y,z)  du^2 - 2 du dv + dy^2 + dz^2,
	\ee
corresponding to a $pp$-wave metric, where $u$ and $v$ are null coordinates. 
For this metric  the nonvanishing components of the Riemann tensor are,
	\be
	\label{Riemann}
		R_{uyuy} = \frac12 F_{yy},\; R_{uyuz} = \frac12 F_{yz},\; R_{uzuz} = \frac12 F_{zz},
	\ee
and those obtained by using the symmetries of $R_{\mu\nu\alpha\beta}$.
In (\ref{Riemann}) we introduced notations $F_{ij} \equiv \partial^{2}F/\partial x^i\partial x^j$.
Therefore to insure that the metric (\ref{metric}) describes a non-trivial solution,
one or more second derivative of $F(u,y,z)$ with respect to $y$ or/and $z$ must be nonvanishing.
For the ansatz (\ref{metric}) the Ricci scalar is identically zero,
$
	R=0.
$
The only nonvanishing component of the Ricci tensor (and therefore of the Einstein tensor)
is the $uu$ component,
	\be
		R_{uu} = \frac12 \left(F_{yy} + F_{zz} \right).\nonumber
	\ee
Since we look for a plane-wave solution, we assume that the scalar field depends only on the null
coordinate $u$,
\be
	\vp = \vp(u). \label{tauansatz}
\ee
For the given ansatz the only possible nonvanishing component of the current is the $u$-component.
By substituting (\ref{tauansatz}) and (\ref{metric}) into the expressions for the currents
(\ref{J2}), (\ref{J3}), (\ref{J4}) and (\ref{J5}) we obtain,
	\be
		J^{(2)}_{\mu}   = K_{X}^{}\vp_{,\mu}, \; J_{\mu}^{(3)}   =  J^{(4)}_{\mu} = J^{(5)}_{\mu} = 0,
	\ee
where $K_X \equiv dK(X)/dX$.
The fact that most of the terms drop out from the expression for the current,
can be seen as follows.
Introducing the notation $k_\alpha \equiv \nabla_\alpha\vp$, we note that the only nonvanishing components of $k_\alpha$ and 
its covariant derivative are  $k_u$ and $k_{u;u}$, correspondingly. 
Having in mind that $g^{uu}=0$, we can see that 
a term containing more than one $k_\alpha$ and only one free (uncontructed) index 
is automatically vanishing. Similarly, a contraction of $k_\alpha$ with the Riemann tensor gives zero for the plane wave ansatz,
and  the scalar curvature is also zero, $R=0$.
Now we can use a key property of the Galileons: the fact that for the higher-order  Lagrangians
free functions of the standard kinetic term $X$ are multiplied by (one or several) $k_{\alpha;\beta}$, by a curvature tensor or both.
Thanks to this property, the currents corresponding to the higher order Galileons contain either more than one $k_\alpha$, 
or contractions of $k_\alpha$ with a curvature tensor. 
In particular, $J^{(3)}$ is of the form $k\nabla k$,
 $J^{(4)} \sim (\nabla k)^2 k $ and   $J^{(5)} \sim (\nabla k)^3 k $. 
Note that $J^{(4)}$ and $J^{(5)}$ also contain terms, involving curvature.
All these terms, however, are vanishing for the plane wave ansatz, for the reason given above.
Therefore only the $k$-essence part of the Lagrangian, $\L^{(2)}$,
gives a non-trivial contribution to the current for the ansatz  (\ref{metric}) and  (\ref{tauansatz}).
Then it is not difficult to check that the scalar field equation of motion, Eq.~(\ref{eomJ}), is satisfied automatically.
Thus any function $\vp(u)$ and metric (\ref{metric}) is a solution of equations of motion for the scalar field.

We need also to make sure that the equations of motion for the metric are satisfied.
Let us calculate the energy-momentum tensor for the given ansatz.
We assume $K(0)=0$ to exclude the cosmological term.
Most of the terms drop out from energy-momentum tensor when the ansatz (\ref{metric}) and (\ref{tauansatz})
is substituted. 
This can be seen by using similar arguments we used above, for the current.
Indeed, it is not difficult to see that only nontrivial parts of the energy-momentum are those proportional to 
 $k_\mu k_\nu$, $ g_{\mu\nu}$ and $ R_{\mu\nu}$ (or, equivalently $G_{\mu\nu}$), with coefficients depending on $X$.
 All other terms vanish, because they contain scalar products of  $k_\mu$ (or its derivative) either with itself or with curvature tensors.
 One can associate the non-vanishing terms with the 
 $k$-essence Lagrangian, giving contributions  $\propto k_\mu k_\nu$,
    \be
    	\label{Ttt1}
        T_{uu}^{(2)}  =  K_{X}(0)\left(\vp'_u\right)^2.
    \ee
 and with $\L^{(4)}$ Lagrangian, giving $R_{\mu\nu}$ contribution to the metric equations of motion,
    \be
    	\label{Ttt2}
        T_{uu}^{(4)}  =   -2 G^{(4)}(0) R_{uu}.
    \ee
In (\ref{Ttt1}) we defined $\vp'_u \equiv d\vp/du$.
The non-trivial component of the metric equation of motion takes the form,
	\be
		\label{F}
		F_{yy} +F_{zz}  =  \kappa \left(\vp'_u\right)^2,
	\ee
where 
\be
\label{kappa}
\kappa \equiv \frac{K_X(0)}{G^{(4)}(0)}=\text{const}.
\ee
Eq.~(\ref{F}) is a two-dimensional (in $y$ and $z$ coordinates) Poisson equation with a constant source.
The general solution of (\ref{F}) contains the homogenous and a particular solutions.
A particular solution for $F$ can be easily found from by integration of (\ref{F}),
	\be
	\label{Fsol}
		F_\vp(u,y) = \frac12 \kappa \left(y\,\vp'_u\right)^2.
	\ee
where we omitted the constant term and the term linear in $u$, since they can be removed by  an appropriate coordinate 
transformation\footnote{The fact that a solution of the form $\sim C_0 +  C_1 u$ with $C_0$ and $C_1$ constants, 
is a pure gauge can be also seen by from~(\ref{Riemann}): such a solution does not contribute to the Riemann tensor.}. 
Note that in three dimensions, in the context of canonical scalar field minimally coupled to gravity without self-interacting potential, 
such solution has been found in \cite{AyonBeato:2005bm}.
The subscript $\vp$ in (\ref{Fsol}) implies that this solution --- a particular solution of the partial differential equation~(\ref{F}) ---
depends on the form of the scalar field profile.
The homogeneous solution of (\ref{F}) satisfies the Laplace equation,
	\be
		\label{FL}
		F_{yy} +F_{zz}  =  0,
	\ee
and can be written as a series,
	\be
	\label{Fh}
		F_g(u,y,z) = \sum f^{(i)}(u) w^{(i)} (y,z).
	\ee
The solution $F_g$ corresponds to a free gravitational wave and independent of $\vp(u)$.
Functions $f^{(i)}$ are arbitrary, while each $w^{(i)}$ in (\ref{Fh}) satisfies the two-dimensional Laplace equation,
	\be
		\label{w}
		w^{(i)}_{yy} +w^{(i)}_{zz}  =  0,\nonumber
	\ee
	where $w^{(i)}_{ij} \equiv \partial^2 w^{(i)}/\partial x^i \partial x^j$.
An important class of solutions for $F_g$ can be written as,
	\be
		\label{Fg}
		F_g(u,y,z) = a(u)(y^2 - z^2) + 2 b(u) yz,
	\ee
where $a$ and $b$ are arbitrary smooth functions.
The metric (\ref{metric}) with $F$ given by~(\ref{Fg}) coincides with the plane gravitational wave in General Relativity.
Finally, the full solution can be written as,
	\be
	\label{fullsol}
	\begin{aligned}
		\vp  & = \vp(u), \\
		ds^2 & = -\left[\frac{\kappa}{2} \left(y\,\frac{d\vp}{du}\right)^2+ F_g(u,y,z) \right] du^2\\
		& -2 du dv + dy^2 + dz^2,
	\end{aligned}
	\ee
where $\vp$ is an arbitrary smooth function of $u$,
and $F_g$ is a solution of (\ref{FL}) and $\kappa$ is given by (\ref{kappa}).

\section{Summary}\label{conclusion}
In this paper we found an exact plane wave solution in the most general scalar-tensor Horndeski
(or generalized Galileon) theory possessing shift-symmetric Lagrangian.
The solution, Eq.~(\ref{fullsol}), is given in terms of an arbitrary smooth function of a light coordinate, $\vp = \vp(u)$,
parametrizing the scalar field profile;
and by a function $F_g(u,y,z)$, which satisfies the homogeneous Laplace equation~(\ref{FL}).
The solution for $F(u,y,z)$, contains two pieces.
One part, $F_\vp$, describes the backreaction of the metric due to the presence of the scalar field,
and it is given in terms of the scalar field profile, Eq.~(\ref{Fsol}).
The second piece, $F_g$, does not depend on $\vp$ and represents a free non-linear gravitational wave,
and it satisfies the Laplace equation (\ref{FL}).
An important example of a free gravitation wave contains two free functions $a(u)$ and $b(u)$, Eq.~(\ref{Fg}),
corresponding to two different polarizations.

A noticeable result of our analysis is that for the ansatz (\ref{metric}) and (\ref{tauansatz})
the conserved current  is not identically zero only for the part of the Galileon Lagrangian,
corresponding to the pure $k$-essence theory, Eq.~(\ref{L2}).
The reason is that the higher order Galileon Lagrangians contain either additional $\nabla\nabla\vp$ or a curvature tensor in the action.
Due to this property, the currents for higher-order Galileon Lagrangians involve
contractions of $\nabla\vp$, $\nabla\nabla\vp$ or/and curvature tensor (unlike the $k$-essence part).
However, contraction of indexes for these tensors gives zero, thanks to the plane wave ansatz.

It is also worth to mention that only few terms in the Galileon energy-momentum tensor contribute to the equation of motion for plane wave ansatz. 
This is due to the same reason that most terms in the Galileon current vanish. 
Namely, the energy momentum tensor for 
higher-order Galileons contain contractions of $\nabla\vp$, $\nabla\nabla\vp$ and curvature tensors,
which vanish for the plane wave ansatz. Only few terms survive that do not contain such combinations: 
the terms coming from the $k$-essence part of the Lagrangian and a piece of the $\L_4$-term.

This means, in particular, that $\L_3$ and $\L_5$ terms do not contribute to the 
energy carried by a plane wave. 
This is not a general property of these Galileon terms, since, for example the Galileon term 
$\L_4$ does contribute to the energy flux in the case of the accretion onto a black hole~\cite{Babichev:2010kj}.

When the scalar field has a trivial configuration, $\vp = \text{const}$, our solution reduces to a General Relativity solution for a free gravitational wave.
In particular, the plane gravitational wave, Eq.~(\ref{Fg}), is a solution for the shift-symmetric Horndeski theory.

On the other hand, in the limit when the metric becomes non-dynamical, we reproduce solutions for certain non-linear scalar field theories known before.
In particular, when the full Horndeski Lagrangian contains only $\L_2$ term,
our solution trivially reduces to a wave solution for the pure $k$-essence~\cite{Babichev:2007dw}.
When the non-covariant version of the DGP-like Galileon is considered (in flat metric), with $K \propto X$, $G^{(3)} \propto X $, $\L_4 = \L_5 =0$,
one easily recovers a ``soliton''-like solution for Galileon, which was studied in \cite{Evslin:2011vh,Evslin:2011rj}.

As we can see, the solution for the generalized Galileon only contains the $k$-essence part 
of the full action and the function of the standard kinetic term $G^{(4)}$ evaluated at $X=0$.
Therefore, in some sense, the higher-order Galileon is decoupled from the solution. 
Is there any difference at all between the plane wave solutions for the Galileon and the pure $k$-essence?
On the level of a background solution there is no difference, apart from the constant $G^{(4)}(0)$, appearing in the solution for the Galileon.
However, the perturbations propagate differently for the Galileon and the pure $k$-essence theories.
The higher-order Galileon terms, although not contributing to the background solution for the plane wave, 
affect the action for perturbations. This means, in particular, that the causal structure 
and the stability properties are different for the Galileon and the  $k$-essence theories.

There are open question left for future work. 
First of all, the stability of plane-wave solutions is to be studied. There are several types of instabilities,
which may arise: ghosts, gradient instability or tachyon instability.
One can notice, for example, that the sign of $K_X$, which controls the presence of a ghost in the $k$-essence theory,
does not seem to affect the existance of the solution (\ref{fullsol}). 
Therefore, a separate analysis of the stability of a plane wave is required.
Since the scalar and the gravity modes are kinetically mixed,
the stability analysis promises to be rather challenging.
Another interesting problem to investigate
is a possible formation of caustics in the plane-wave solution.
Since the perturbations of the Galileon on top of the plane wave background
may travel with the speed, exceeding the speed of light
(while the background solution travels with the speed of light), the perturbations tend to accumulate
at the front of the wave. This may be an indication of caustic formation.
Collision of Galileon plane waves is another interesting topic for future work.

\begin{acknowledgments}
It is a pleasure to thank Mokhtar Hassaine for very helpful discussions and critical reading of the manuscript, 
and Gilles~Esposito-Far\`ese for interesting discussions.
 \end{acknowledgments}

\appendix
\section{Lagrangian $\L_5$}
\label{appendix}
Here, for completeness, we list the expressions for the current and the energy-momentum tensor corresponding to the $\L_5$-term~(\ref{L5}). 
These expressions, e.g., can be found from corresponding formulae in Ref.~\cite{Gao:2011qe}, by requiring the shift symmetry, $\vp\to \vp +$const.
For the conserved current one finds,
\begin{widetext}
\begin{eqnarray}
	J^{(5)\mu} & = & -\left[6G_{\rho\sigma}\nabla^{\rho}\nabla^{\sigma}\vp G_{,X}^{(5)} - \left(\left(\square\vp\right)^{3}-3\square\vp\left(\nabla\nabla\vp\right)^{2}+2\left(\nabla\nabla\vp\right)^{3}\right)G_{,XX}^{(3)}\right]\nabla^{\mu}\vp
\nonumber
\\
 &  & +\nabla_{\nu}\left[G_{,X}^{(3)}\left(3\left(\Box\vp\right)^{2}g^{\mu\nu}-3g^{\mu\nu}\left(\nabla\nabla\vp\right)^{2}-6\Box\vp\nabla^{\mu}\nabla^{\nu}\vp+6\nabla^{\mu}\nabla_{\lambda}\vp\nabla^{\lambda}\nabla^{\nu}\vp\right)\right]
 -6G^{\mu\nu}\nabla_{\nu}G^{(5)}.
 \label{J5}
\end{eqnarray}
The energy-momentum tensor reads,
    \be
    \begin{aligned}
    \label{T5}
        T_{\mu\nu}^{(5)} & =  C_{1}g_{\mu\nu}+C_{2}\nabla_{\mu}\vp\nabla_{\nu}\vp+C_{3}\nabla_{\mu}X\nabla_{\nu}X+C_{4}\nabla_{(\mu}\vp\nabla_{\nu)}X+C_{5}\nabla_{\mu}\nabla_{\nu}\vp+C_{6}\nabla_{\beta}X\nabla^{\beta}\nabla_{(\nu}\vp\nabla_{\mu)}\vp\\
         &   +C_{7}\text{ }\nabla^{\lambda}X\nabla_{\lambda}\nabla_{\beta}\vp\nabla^{\beta}\nabla_{(\mu}\vp\nabla_{\nu)}\vp+C_{8}\nabla^{\alpha}\nabla_{\mu}\vp\nabla_{\nu}\nabla_{\alpha}\vp+C_{9}\nabla_{\text{\ensuremath{\beta}}}X\nabla_{(\nu}X\nabla^{\text{\ensuremath{\beta}}}\nabla_{\mu)}\vp \\
         &   +C_{10}\nabla_{\beta}\nabla_{\alpha}\vp\nabla^{\alpha}\nabla_{\mu}\vp\nabla^{\beta}\nabla_{\nu}\vp+\tau_{\mu\nu}^{(5)},
    	\end{aligned}
        \ee
where,
    \be
    \label{C}
    \begin{aligned}
	C_{1} & =
		 -3G_{,X}^{(5)}\Big[\frac{2}{3}\left(\left(\Box\vp\right)^{3}-3\square\vp\left(\nabla\nabla\vp\right){}^{2}+2\left(\nabla\nabla\vp\right)^{3}\right)
		 -2\square\vp R_{\rho\sigma}\nabla^{\rho}\vp\nabla^{\sigma}\vp \\
		 & -4R_{\rho\sigma}\nabla^{\rho}X\nabla^{\sigma}\vp+R\nabla_{\rho}X\nabla^{\rho}\vp
		 +2R_{\rho\lambda\sigma\tau}\nabla^{\rho}\vp\nabla^{\sigma}\vp\nabla^{\lambda}\nabla^{\tau}\vp\Big] \\
 		&   -G_{,XX}^{(5)}\left[3\left(\left(\square\vp\right)^{2}-\left(\nabla\nabla\vp\right){}^{2}\right)\nabla_{\lambda}X\nabla^{\lambda}\vp
		-6\nabla^{\rho}X\nabla^{\sigma}X\nabla_{\rho}\nabla_{\sigma}\vp+6\square\vp\nabla^{\rho}X\nabla_{\rho}X\right],\\
		C_{2} & =  -6G_{,X}^{(5)}G_{\rho\sigma}\nabla^{\rho}\nabla^{\sigma}\vp
		 	+G_{,XX}^{(3)}\left[\left(\square\vp\right)^{3}-3\left(\nabla\nabla\vp\right)^{2}\square\vp+2\left(\nabla\nabla\vp\right)^{3}\right],\\
		C_{3}  & =  6G_{,XX}^{(5)}\square\vp,\quad
		C_{4} =  6G_{,X}^{(5)}R+6G_{,XX}^{(5)}\left[\left(\square\vp\right)^{2}-\left(\nabla\nabla\vp\right)^{2}\right],\\
		C_{5} & =  -6G_{,X}^{(5)}\left[\left(\nabla\nabla\vp\right)^{2}-\left(\square\vp\right)^{2}+R_{\rho\sigma}\nabla^{\rho}\vp\nabla^{\sigma}\vp\right]
		 	+6G_{,XX}^{(5)}\left(\nabla_{\rho}X\nabla^{\rho}X+\square\vp\nabla_{\rho}X\nabla^{\rho}\vp\right),\\
		C_{6} & =  -12G_{,XX}^{(5)}\square\vp,\quad		C_{7}  =  12G_{,XX}^{(5)},\quad
		C_{8}  =  -12G_{,X}^{(5)}\square\vp-6G_{,XX}^{(5)}\nabla_{a}X\nabla^{a}\vp,\\
		C_{9} & =  -12G_{,XX}^{(5)},\quad		C_{10}  =  12G_{,X}^{(5)},\\
		\tau_{\mu\nu}^{(5)} & =
		 -6G_{,X}^{(3)}\Big\{2\left[\left(\square\vp\nabla^{\rho}\vp+\nabla^{\rho}X\right)R_{\rho(\mu}
		 +\nabla^{\sigma}\nabla^{\rho}\vp\nabla^{\lambda}\vp R_{\lambda\rho\sigma(\mu}
		 -R_{\rho\sigma}\nabla^{\rho}\vp\nabla^{\sigma}\nabla_{(\mu}\vp\right]\nabla_{\nu)}\vp\\
 		& -\nabla_{\rho}X\nabla^{\rho}\vp R_{\mu\nu}+2\nabla^{\rho}\vp R_{\rho(\mu}\nabla_{\nu)}X
		-R_{\rho(\mu\nu)\sigma}\nabla^{\rho}\vp\left(\square\vp\nabla^{\sigma}\vp+2\nabla^{\sigma}X\right)
		-2\nabla^{\rho}\vp\nabla^{\lambda}\vp R_{\lambda\sigma\rho(\mu}\nabla^{\sigma}\nabla_{\nu)}\vp\Big\}.
	\end{aligned}
	\ee
\end{widetext}

\end{document}